# Necessary criterion of choosing the energy-momentum tensor and the Lagrange formalism

Yurii A. Spirichev

The State Atomic Energy Corporation ROSATOM, "Research and Design Institute of Radio-Electronic Engineering" - branch of Federal Scientific-Production Center "Production Association "Start" named after Michael V.Protsenko", Zarechny, Penza region, Russia

E-mail: yurii.spirichev@mail.ru

**Abstract**

It is shown that the necessary criterion for choosing the energy-momentum tensor of a physical system is the form of its linear invariant, which should be the lagrangian of this physical system. Examples of energy-momentum tensors that meet this criterion are given. For an electromagnetic field in vacuum, the linear invariant of the energy-momentum tensor must be the canonical lagrangian or the quadratic invariant of the electromagnetic field tensor. A mathematical method for obtaining the lagrangian that complements the Lagrange formalism is proposed.

## 1. Introduction

The energy-momentum tensors (EMT) play a key role in the description of physical processes, since they are followed by the equations of the laws of conservation of energy and momentum. The equations of conservation of momentum are simultaneously the equations of motion. In this regard, the criteria for assessing the correctness of the EMT play an important role in its construction or selection. Such a criterion can be the correspondence of the structure of the considered EMT to the structure of the canonical EMT [1]. However, this criterion is not unambiguous, as it can correspond to a number of EMT. Another frequently used criterion for selecting EMT is its symmetry. However, this criterion is not unambiguous either. An example of this is the EMT of Minkowski and Abraham and the well-known long discussion about their correctness [2]. Consequently, it can be concluded that nowadays there is no unambiguous and necessary criterion for assessing the correctness of EMT in electrodynamics. This leads to the existence of many competing EMT and discussions about their correctness.

The principle of least action occupies an important position in physics. The invariance of an action with respect to transformations of the Poincaré group using the Lagrange formalism allows one to obtain equations of motion and equations for the conservation of energy and momentum of a physical system. These equations also follow from the EMT, which indicates that the EMT is closely related to the action of the physical system. Therefore, one of the ways to obtain EMT is to use the Lagrange formalism [1]. To use it, it is necessary to know the density of the Lagrange function or the lagrangian of a physical system. However, there are no rigorous mathematical methods in physics for obtaining the lagrangian and it is obtained empirically, based on general considerations [3-6]. As the lagrangian is an invariant and is associated with the energy of the physical system, and the other invariant associated with energy is the track of EMT, it is natural to assume that the trace of EMT is also the lagrangian of the physical system.

The purpose of this article is to prove that the necessary criterion for choosing an EMT, providing an unambiguous assessment of its correctness, is the kind of the trace of EMT of a physical system, which simultaneously must be its lagrangian.

### 2 Necessary criterion for selecting EMT

An important characteristic of EMT is its linear invariant, which is the sum of its diagonal components.

$$I = T_{00} + T_{11} + T_{22} + T_{33} \quad (1)$$

Canonical EMT has a general form [1]

$$T_{\mu\nu} = \begin{bmatrix} W & i\frac{1}{c}\mathbf{S} \\ ic\cdot\mathbf{g} & t_{ik} \end{bmatrix} \quad (\mu, \nu = 0, 1, 2, 3; \ i, k = 1, 2, 3) \quad (2)$$

where $W$ – energy density; $\mathbf{S}$ – energy flux density; $\mathbf{g}$ – momentum density; $t_{ik}$ – momentum flux density tensor (stress tensor). The trace of this tensor, which is its linear invariant, has the form:

$$I = W + t_{ii} \quad (3)$$

For the mechanical energy-momentum tensor, this invariant has the form:

$$I = m\cdot c^2 - \mathbf{v}\cdot\mathbf{p} \quad (4)$$

where m - is the mass density; $\mathbf{p}$ - is the density of the mechanical momentum of the medium; $\mathbf{v}$ - is the velocity of the medium particles. This invariant, after transition from mass density to mass, corresponds to the canonical density expression of lagrangian known in mechanics [1] for a free particle.

EMT of interaction of the electromagnetic field (EMF) with electric charges is known [7]:

$$T^E_{\mu\nu} = \mathbf{A}_\mu \otimes \mathbf{J}_\nu = \begin{pmatrix} \rho\cdot\varphi & \frac{1}{c}i\cdot\varphi\cdot J_x & \frac{1}{c}i\cdot\varphi\cdot J_y & \frac{1}{c}i\cdot\varphi\cdot J_z \\ i\cdot c\cdot\rho\cdot A_x & -A_x\cdot J_x & -A_x\cdot J_y & -A_x\cdot J_z \\ i\cdot c\cdot\rho\cdot A_y & -A_y\cdot J_x & -A_y\cdot J_y & -A_y\cdot J_z \\ i\cdot c\cdot\rho\cdot A_z & -A_z\cdot J_x & -A_z\cdot J_y & -A_z\cdot J_z \end{pmatrix} \quad (5)$$

where $\mathbf{A}_\mu(\varphi/c, i\cdot\mathbf{A})$ - electromagnetic potential of electromagnetic field, $\varphi$ and $\mathbf{A}$ - scalar and vector potentials of electromagnetic field; $\mathbf{J}_\nu(c\cdot\rho, i\cdot\mathbf{J})$ - four-dimensional current density, $\rho$ and J - charge density and conduction current density vector. The trace of this EMT or its linear invariant has the form:

$$I = \rho \cdot \varphi - \mathbf{A} \cdot \mathbf{J} \qquad (6)$$

This invariant is known as the generalized energy density of electromagnetic interaction of electromagnetic fields with electric charges [8] and it is also the lagrangian of this interaction.

EMT of interaction of EMF with dielectric medium is known [9]:

$$T_{\mu\nu} = \begin{bmatrix} \mathbf{E} \cdot \mathbf{D} & i\frac{1}{c}(\mathbf{E} \times \mathbf{H}) \\ ic \cdot (\mathbf{D} \times \mathbf{B}) & E_i D_k + B_i H_k - \delta_{ik}(\mathbf{B} \cdot \mathbf{H}) \end{bmatrix} \qquad (7)$$

The linear invariant of this EMT has the form:

$$I = \mathbf{E} \cdot \mathbf{D} - \mathbf{B} \cdot \mathbf{H} \qquad (8)$$

For EMF in vacuum, EMT (7) takes the form:

$$T_{\mu\nu}^V = \begin{bmatrix} \varepsilon_0 \cdot \mathbf{E}^2 & i \cdot (\mathbf{E} \times \mathbf{B})/c \cdot \mu_0 \\ i \cdot c \cdot \varepsilon_0 \cdot (\mathbf{E} \times \mathbf{B}) & \varepsilon_0 \cdot E_i E_k + B_i B_k / \mu_0 - \delta_{ik} \mathbf{B}^2 / \mu_0 \end{bmatrix} \qquad (9)$$

The linear invariant of this EMT is has form:

$$I = \mathbf{E}^2 / c^2 - \mathbf{B}^2 \qquad (10)$$

This linear invariant of EMT (9) is the canonical lagrangian of EMF in vacuum [6] and, simultaneously, a quadratic invariant of the antisymmetric tensor of EMF $F_{\mu\nu}$:

$$I = F^{\mu\nu} F_{\mu\nu} = \mathbf{E}^2 / c^2 - \mathbf{B}^2 \qquad (11)$$

Since this quadratic invariant (11), containing the EMF energy, is the only invariant, then the linear invariant of the EMT (10) is also the only possible invariant. From this, it follows that the EMTs (7) and (9) are correct. Thus, a necessary criterion for the correctness of any EMT for an EMF in a medium is the equality of its linear invariant (1), in the "vacuum" approximation, to the invariant (11) of the EMF.

Let us consider the linear invariant of the EMT of Minkowski. Its components are equal:

$$T_{00} = (\mathbf{E} \cdot \mathbf{D} + \mathbf{H} \cdot \mathbf{B})/2$$
$$T_{11} = -(E_y \cdot D_y + H_y \cdot B_y + E_z \cdot D_z + H_z \cdot B_z)/4 \qquad (12)$$
$$T_{22} = -(E_x \cdot D_x + H_x \cdot B_x + E_z \cdot D_z + H_z \cdot B_z)/4$$
$$T_{33} = -(E_x \cdot D_x + H_x \cdot B_x + E_y \cdot D_y + H_y \cdot B_y)/4$$

If we consider the vacuum as a medium with relative dielectric and magnetic permeability's equal to one, then this invariant of the EMT will describe the energy density of the electromagnetic field in a vacuum.

$$T_{00} = (\mathbf{E}^2 / c^2 + \mathbf{B}^2)/2$$
$$T_{11} = -(E_y^2 / c^2 + B_y^2 + E_z^2 / c^2 + B_z^2)/4 \qquad (13)$$

$$T_{22} = -(E_x^2/c^2 + B_x^2 + E_z^2/c^2 + B_z^2)/4$$

$$T_{33} = -(E_x^2/c^2 + B_x^2 + E_y^2/c^2 + B_y^2)/4$$

Substituting the components of the EMT of Minkowski in the "vacuum" approximation (13) into the invariant (1), we obtain:

$$I = (\mathbf{E}^2/c^2 + \mathbf{B}^2)/2 - (E_y^2/c^2 + B_y^2 + E_z^2/c^2 + B_z^2)/4 - (E_x^2/c^2 + B_x^2 + E_z^2/c^2 + B_z^2)/4 -$$
$$- (E_x^2/c^2 + B_x^2 + E_y^2/c^2 + B_y^2)/4 = 0$$

From this expression it follows that the Minkowski EMT does not satisfy the necessary correctness criterion, since its linear invariant in the "vacuum" approximation is not equal to the EMF invariant (11) and the canonical lagrangian of the EMF in a vacuum. Similarly, you can check other known EMT. Thus, EMT (7) is correct for an EMF in dielectric medium.

### 3 Energy-momentum tensor and Lagrange formalism

Lagrange formalism plays a key role in field theory. However, there is a problem in this formalism as there are no rigorous mathematical methods for obtaining the lagrangian. Therefore, it is obtained by the method of construction (selection, design) on the basis of general considerations or first, in one way or another, the equations of motion of the physical system are obtained, and then the lagrangian is adjusted to these equations. With this method, the theoretical results obtained are not always unambiguous, since the choice of the lagrangian largely depends on the views and imagination of the researcher. From the examples given in Section 2, it was concluded that the lagrangians of physical systems are linear invariants or traces of the EMT of these systems. Thus, obtaining the lagrangian from the EMT for further application in the Lagrange formalism is a mathematically rigorous method. The lagrangian obtained by this method can be written in the form $L = \delta^{\nu\mu} T_{\nu\mu}$, where $\delta^{\nu\mu}$ - is the Kronecker symbol, $T_{\nu\mu}$ - EMT. The sequence of actions is the following: firstly, we obtain the field tensor $F_{\nu\mu}$, then from the field tensor using the method described in [9], we get the EMT $T_{\nu\mu} = F_{\nu\eta} F_{\eta\mu}$, then take its trace, which in the form of lagrangian, is used in the Lagrange formalism. Thus, the proposed method can be considered as an addition to the Lagrange formalism, which provides a mathematically rigorous way of obtaining of the lagrangian of the physical field.

The equations of motion of the field follow from the tensor of a physical field, and the equations of energy conservation, energy flux density and field momentum follow from EMT. These equations can also be obtained using the Lagrange formalism, if you know the lagrangian. Thus, the equations of motion obtained using the Lagrange formalism duplicate the previously obtained equations from the field tensor and the EMT. Consequently, the proposed method of obtaining the lagrangian and the formalism Lagrange form a closed chain of mathematical transformations and provide mutual verification of the correctness of the components of this chain (field tensor, equations of motion, TEI, equations of energy and momentum conservation, the lagrangian). The structure of the mathematical operations of such a formalism for a field in a vacuum is shown in the figure.

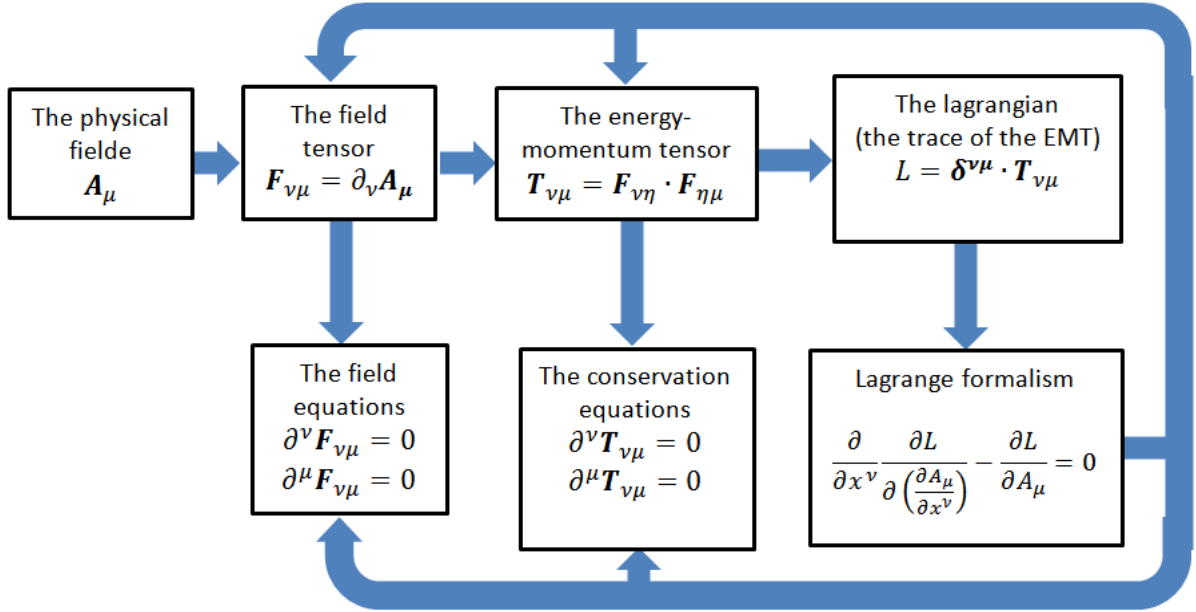

The structure of a formalism

## 4. Conclusion

The examples of Section 2 show that the trace EMT, which is a linear invariant of the EMT, is also the lagrangian of the physical system that this EMT describes. Thus, the equality of the lagrangian to the EMT is a necessary criterion for the validity of the EMT, more precisely, the correspondence between the lagrangian and the EMT of the physical system.

For EMF in vacuum, the linear invariant of the EMT is a quadratic invariant of EMF or its canonical lagrangian. This follows from the fact that the diagonal components of the EMT represent the energy of the electromagnetic field, and the sum of these components, i.e. the trace of a tensor, is an invariant. But there is only a single invariant representing the energy of the electromagnetic field, it is the canonical invariant of the electromagnetic field or its lagrangian.

The energy-momentum tensors corresponding to this criterion are given. The proposed method allows, firstly, to mathematically strictly obtain the lagrangians of physical systems, secondly, on the basis of the Lagrange formalism, to obtain a closed formalism that provides verification of the correctness and consistency of the original, intermediate and final theoretical research results.

**References**
1. Landau L D, Lifshits E M. *The Classical Theory of Fields* (Oxford: Pergamon Press, 1983).

2. Kirk T. McDonald. Bibliography on the Abraham-Minkowski debate (Feb.17, 2015, updated September 29, 2017).
3. Y. Aharonov and D. Bohm, Phys. Rev. 130 (1963) 1625.
4. Erez Raicher, Shalom Eliezer and Arie Zigler. The Lagrangian Formulation of Strong-Field QED in a Plasma. arXiv: 1312.3088.
5. D. M. Gitman, A. E. Shabad, and A.A. Shishmarev. Moving point charge as a soliton in nonlinear electrodynamics. arXiv: 1509.06401.
6. Srivatsan Rajagopal, Ajit Kumar. Quantization of B-I electrodynamics and B-I modified gravity using Faddeev-Popov gauge-fixing procedure. arXiv: 1305.0083.
7. Yurii A. Spirichev, *About the Abraham force in a conducting medium*. arXiv: 1707.08642.
8. Zommerfeld A., Electrodinamix, M.; 1958.
9. Yurii A. Spirichev, *A new form of the energy-momentum tensor of the interaction of an electromagnetic field with a non-conducting medium. The wave equations. The electromagnetic forces*, arXiv: 1704.03815.